\documentclass[12pt]{article}
\usepackage{helvet, graphicx}
\usepackage{cite}
\usepackage{hyperref}
\usepackage{gensymb}
\usepackage[usenames]{color}
\usepackage{amsmath,amssymb}
\usepackage{bm}
\usepackage{braket}
\usepackage[margin=0.75in]{geometry}
\usepackage[T1]{fontenc}
\usepackage{caption}
\usepackage[font=small]{caption} 
\usepackage[square,numbers]{natbib}
\usepackage{setspace}

\newcommand{\er}{Er$^{3+}$ }
\newcommand{\upg}{\uparrow_\text{g}}
\newcommand{\downg}{\downarrow_\text{g}}
\newcommand{\upe}{\uparrow_\text{e}}

\newcommand{\td}{T_\text{1,dark}}

\title{Optical quantum nondemolition measurement of a solid-state spin without a cycling transition}
\author{Mouktik Raha, Songtao Chen, Christopher M.~Phenicie, Salim Ourari,\\ Alan M.~Dibos\thanks{Present address: Nanoscience and Technology Division, Argonne National Laboratory, Argonne, IL 60439, USA}, Jeff D.~Thompson\thanks{jdthompson@princeton.edu}}
\date{%
    \centering
    \textit{Department of Electrical Engineering, Princeton University, Princeton, NJ 08544, USA}\\[2ex]%
    \today
}

\begin{document}

\maketitle
\abstract{Optically-interfaced spins in the solid state are a promising platform for quantum technologies. A crucial component of these systems is high-fidelity, projective measurement of the spin state. In previous work with laser-cooled atoms and ions, and solid-state defects, this has been accomplished using fluorescence on an optical cycling transition; however, cycling transitions are not ubiquitous. In this work, we demonstrate that modifying the electromagnetic environment using an optical cavity can induce a cycling transition in a solid-state atomic defect. By coupling a single Erbium ion defect to a telecom-wavelength silicon nanophotonic device, we enhance the cyclicity of its optical transition by a factor of more than 100, which enables single-shot quantum nondemolition readout of the ion's spin with 94.6\% fidelity. We use this readout to probe coherent dynamics and relaxation of the spin. This approach will enable quantum technologies based on a much broader range of atomic defects.}\\

Atomic and atom-like defects in the solid state provide an optical interface to individual electronic and nuclear spin qubits \cite{Awschalom:2018ic}, and are used for a variety of quantum technologies. As sensors, they can probe temperature and magnetic and electric fields with nanoscale spatial resolution \cite{Kucsko:2013cz,Maletinsky:2012ge,Dolde:2011hi}. In quantum networks, spin-photon entanglement \cite{Togan:2010hia,DeGreve:2012gea,Gao:2012ir} has enabled deterministic entanglement of remote spins \cite{Bernien:2013kj}. Defect spins have also been used to demonstrate key components of quantum information processors, including quantum error correction \cite{Taminiau:2014chb} and 10-qubit quantum registers with multi-qubit gates \cite{bradley201910}.

These works primarily leverage the well-studied nitrogen vacancy (NV) center in diamond. However, a much broader range of defects exists that may be advantageous for particular applications. For example, the SiV$^-$ \cite{Sipahigil:2016hy} and SiV$^0$ \cite{Rose:2018hp} color centers in diamond are promising for quantum networks because of their low spectral diffusion, while color centers in silicon carbide \cite{Christle:2014kg} may be easier to integrate with nanoscale devices. Rare earth ions are another family of defects that can offer long spin coherence \cite{Zhong:2014bw} and narrow, stable optical transitions (in the telecom band {for} the case of \er\!\!) \cite{Bottger:2009ik}, and may be doped into a variety of host crystals. Several recent works have begun to probe individual rare earth ions \cite{Yin:2013jva,Siyushev:2014iea,Utikal:2014bda, Nakamura:2014gta,Dibos:2018hh,Zhong:2018ie}, using an optical cavity to overcome their  low intrinsic photon emission rates \cite{Dibos:2018hh,Zhong:2018ie}.

A key capability for atomic defects is high-fidelity spin readout using the optical transition \cite{Awschalom:2018ic}. Single-shot optical spin measurements have been achieved in quantum dots \cite{Delteil:2014er} and in the NV \cite{Robledo:2011fs} and SiV$^-$ \cite{Sukachev:2017kv} color centers in diamond by leveraging highly cyclic optical transitions that arise from atomic selection rules. However, cyclic optical transitions are not a universal feature of atomic defects, and are often absent in low-symmetry defects and in the presence of strain \cite{Tamarat:2008ko} or spin-orbit coupling without careful alignment of the magnetic field \cite{Sukachev:2017kv,Delteil:2014er}. Single-shot readout has not been achieved in atomic defects without intrinsic cycling transitions, such as rare earth ions \cite{Bartholomew:2016hs}.

In this work, we demonstrate that tailoring the electromagnetic density of states around an atom with an optical cavity can induce highly cyclic optical transitions in an emitter that is not naturally cyclic. Using a single \er ion in Y$_2$SiO$_5$ (YSO) coupled to a silicon nanophotonic cavity (Fig.~1a), we demonstrate a greater than 100-fold enhancement of the cyclicity: under conditions where the ion alone can scatter fewer than {10} photons before flipping its spin, a cavity-coupled ion can scatter over {1200}. This is sufficient to realize single-shot spin readout with a fidelity of 94.6\%, and to enable continuous, quantum nondemolition measurement of quantum jumps between the ground state spin sublevels. The improvement in the cyclicity arises from selective Purcell enhancement of the spin-conserving optical decay pathway (Fig.~1b), determined primarily by the alignment of the cavity polarization and the spin quantization axis defined by a magnetic field. A small additional enhancement arises from detuning of the spin-non-conserving transitions from the optical cavity, an effect that was recently used to enhance the cyclicity of a quantum dot in a nanophotonic cavity \cite{Sun:2018hx}. This generic technique opens the door to exploiting a much broader range of atomic defects for quantum technology applications, and is a particular advance for individually addressed rare earth ions.

Our experimental approach, following Ref.~\citealp{Dibos:2018hh}, is based on a YSO crystal doped with a low concentration of \er ions placed in close proximity to an optical cavity in a silicon photonic crystal waveguide (Fig.~1a). Assembled devices are placed inside a $^3$He cryostat at 0.54 K with a 3-axis vector magnet. Light is coupled to the cavities using a lensed optical fiber on a 3-axis translation stage. The high quality factor ($6 \times 10^4$) and small mode volume of the cavity, together with the high radiative efficiency of the \er optical transition, enable Purcell enhancement of the \er emission rate by a factor of $P=700$ (Fig.~2a). There are several hundred ions within the mode volume of the cavity, but their optical transitions are inhomogeneously broadened over a several GHz span, such that stable, single ion lines can be clearly isolated (Fig.~1c) \cite{Dibos:2018hh}.

The ground and excited states of the 1.536 $\mu$m optical transition in \er\!\!:YSO are effective spin-1/2 manifolds, which emerge from the lowering of the 16- (14-)fold degeneracy of the $^4$I$_{15/2}$ ground ($^4$I$_{13/2}$ excited) states by the crystal field. In the absence of a magnetic field, the ground and excited states are two-fold degenerate, as required by Kramers' theorem  \cite{abragam2012electron}. This degeneracy is lifted in a small magnetic field, revealing four distinct optical transitions (Fig.~1b,c). Transitions A and B conserve the spin, while C and D flip the spin.

To probe the selection rules of the optical transition, we excite the spin-conserving A and B lines alternately~(Fig.~\ref{fig:2}a). The average fluorescence following the A and B pulses is the same, since the transitions are symmetrically detuned from the cavity and the spin is on average unpolarized from continuous optical pumping by the excitation light. However, the intensity autocorrelation function, $g^{(2)}(n t_\text{rep})$ (where $n$ is the offset in the number of pulses) is anti-bunched for odd-numbered pulse offsets (\emph{i.e.}, A-B correlations) and bunched for even offsets (\emph{i.e.}, A-A or B-B correlations), revealing that only one of the transitions A or B is bright at any given time, depending on the instantaneous spin state (Fig.~2b). Note that the fluorescence after each pulse is integrated before computing the autocorrelation, so $g^{(2)}(n t_\text{rep})$ is only defined for discrete times. Eventually, the spin relaxes and $g^{(2)}$ decays exponentially to 1 after an average of $n_0$ pulses. Under the assumption (to be verified later) that the observed spin relaxation arises primarily from optical pumping between the spin sublevels, we extract the optical transition cyclicity $C = n_0 P_\text{ex}$, where {$P_\text{ex} \approx 1/2$} is the probability to excite the ion in each pulse. This value of $P_\text{ex}$ is assured by using an intense excitation pulse to saturate the ion, and is verified using the independently measured collection efficiency \cite{suppinfo}.

We repeat this measurement with different orientations of the magnetic field, and find that the cyclicity varies by nearly three orders of magnitude (Fig.~2{c,d}), with a maximum value of {$1260 \pm 126$}. This results from the changing orientation of the atomic transition dipole moment with respect to the cavity polarization. It can be captured by a simple model where the decay rates on each transition are proportional to the projection of an associated dipole moment $\vec{d}$ onto the cavity polarization $\vec{\epsilon}$ at the position of the ion. For the spin conserving transition, $\Gamma_\text{AB} \propto \left|\bra{\upe} \vec{\epsilon} \cdot \vec{d}_{||} \ket{\upg} \right|^2$, while $\Gamma_\text{CD}$ is defined analogously with $\vec{d}_{||}$ replaced by $\vec{d}_\perp$. When the magnetic field is rotated, the spin sublevels mix such that $\ket{\uparrow(\varphi',\theta')} = \alpha \ket{\uparrow(\varphi,\theta)} + \beta \ket{\downarrow(\varphi,\theta)}$, with the coefficients $\alpha$, $\beta$ completely specified by the anisotropic $\textbf{g}$ tensor describing the Zeeman shifts (Fig.~1d) \cite{Sun:2008ik}. Together with the time-reversal symmetry properties of the Kramers' doublets, this allows the complete angular dependence of {$C = \Gamma_{AB}/\Gamma_{CD} + 1$} to be described by only two parameters: $\vec{\epsilon} \cdot \vec{d}_{||}$ and $\vec{\epsilon} \cdot \vec{d}_{\perp} $ at a single (arbitrary) reference orientation \cite{suppinfo}. In this model, the role of the cavity is to restrict the decay to a particular polarization, such that the decay rates are determined by a single matrix element $|\vec{\epsilon} \cdot \vec{d}|^2$; in free space, there is no preferred $\vec{\epsilon}$.

Since the dipole matrix elements for \er\!\!:YSO and the cavity field polarization at the position of the atom are not known, we treat $\vec{\epsilon} \cdot \vec{d}_{||}$ and $\vec{\epsilon} \cdot \vec{d}_{\perp}$ as fit parameters. A fit to this model displays excellent agreement with the data, and allows the complete angle dependence of the cyclicity to be extracted from a small number of measurements. While this discussion centers on electric dipole coupling, the \er transition we study has comparable electric and magnetic dipole matrix elements \cite{Dodson:2012dq}, and the predicted magnetic Purcell factor for our structures is similar \cite{Dibos:2018hh}, depending on the precise position of the ion. We show in the Supplementary Information \cite{suppinfo} that the electric and magnetic contributions have the same angular dependence and may be summed into a single term.

To quantify the extent to which the cyclicity is enhanced by the cavity, we study a second ion with lower Purcell factor and then lower it further by detuning the cavity. The cyclicity is observed to decrease roughly linearly with $P$ (Fig.~2e). Based on the dependence of the cyclicity on the cavity detuning for this ion, we estimate that the cyclicity $C_0$ of the ion alone is less than {10} \cite{suppinfo}, such that the enhancement by the cavity is greater than 100. We note that $C_0$ has not been directly measured for \er\!\!:YSO.

Next, we focus on using the cavity-enhanced cyclicity to measure the spin state. Fig.~3a shows a time trace of photons recorded in a single run of the experiment, with telegraph-like switching between $\ket{\downg}$ (where transition B is bright) and $\ket{\upg}$ (where transition A is bright) clearly visible. A continuous estimation of the spin state occupation using a Bayesian estimator applied to the full measurement record \cite{Gammelmark:2014de} shows clearly resolved quantum jumps between these states, demonstrating the quantum nondemolition nature of the measurement. The quantum jumps are driven by optical pumping from the measurement process itself, because of the finite cyclicity.

To demonstrate single-shot measurement of the spin, we use a maximum likelihood (ML) algorithm to estimate the state at time $t$ using photon counts from times $t'>t$. The measurement duration is adaptive: each measurement terminates when a set fidelity threshold or time limit is reached, and a new, independent measurement is begun~\cite{Hume:2007kj}. The outcome of each measurement is shown by the circles in Fig.~3b. The average measurement fidelity estimated by the ML algorithm is 94.6\%, and 91\% of consecutive measurements have the same outcome. The average time to complete a measurement is 20 ms, which corresponds to the average time to detect two photons. The optimum fixed measurement window is 51 ms, resulting in a slower measurement with a lower average fidelity of 91.1\% (Fig.~3c).

Lastly, we apply this spin readout technique to investigate the ground state spin dynamics. We measure the intrinsic spin relaxation rate $\td$ by reducing the optical excitation rate $1/t_\text{rep}$ until the total spin lifetime $T_1 = 1/(\td^{-1} + T^{-1}_{1,\text{op}})$ saturates (Fig.~4a). $T_{1,\text{op}} = C t_\text{rep} /P_\text{ex}$ is the optical pumping time. $\td$ increases with increasing magnetic field strength, in a manner that starkly diverges from the expected $B^{-4}$ behavior of spin-lattice relaxation (Fig.~4b)\cite{abragam2012electron}. One possible explanation is flip-flop interactions with nearby \er ions \cite{car2018optical}, which is consistent with the fact that $T_{1,\text{dark}}$ varies sharply with the magnetic field angle and is different by a factor of 4 between two ions studied~\cite{suppinfo}. In this device, the average separation between magnetically equivalent \er ions is estimated to be 70~nm, such that the dipole-dipole interaction strength is around 1~kHz; the flip-flop rate is likely much slower because of spectral diffusion from nearby $^{89}$Y nuclear spins. The longest relaxation time we observe, $\td = 12.2 \pm 0.4$~s, is the longest electronic spin $T_1$ measured for \er\!\!\!, to the best of our knowledge \cite{Probst:2013hn}.

In Fig.~4c, we demonstrate high-visibility Rabi oscillations between the ground state spin sublevels, driven by a microwave magnetic field applied through a coplanar waveguide. We measure $T_2^* = 125 \pm 5$ ns (in a Ramsey experiment), and $T_2 = 3.3 \pm 0.2\,\mu$s (Hahn echo), consistent with previous measurements of electron spin coherence in solid-state hosts with abundant nuclear spins \cite{Siyushev:2014iea, Petta:2005kn}. Longer coherence times to enable storage of quantum states and the observation of coherent dynamics between interacting \er ions may be achieved using dynamical decoupling. Ultimately, it will be beneficial to use alternative host crystals with lower nuclear spin content; \er incorporation has been demonstrated in several candidates including CaWO$_4$, Si and TiO$_2$ \cite{phenicie2019}.

Our results demonstrate that the optical properties of atomic systems are malleable through control of their local environment. Using a photonic nanostructure, we have achieved more than two orders of magnitude improvement in the emission rate and cyclicity of a single \er ion, and demonstrated single-shot readout of its spin. Realistic improvements in the quality factor of the optical cavity and photon collection efficiency $\eta$ will enable another 20-fold enhancement in emission rate and spin readout with $F>0.99$ in $50 \, \mu$s ($Q = 10^6$ and $\eta = {0.2}$). These results represent a significant step towards realizing quantum networks based on single \er ions. This measurement approach may also be extended to address many closely-spaced \er spins in the same device by exploiting small differences in their optical transition frequencies, providing a foundation for studying strongly interacting spin systems. Finally, this technique will enable a much broader class of atomic defects to be explored for quantum technologies.

We acknowledge helpful conversations with Charles Thiel, Nathalie de Leon and Alp Sipaghil. Support for this research was provided by the National Science Foundation (NSF, EFRI ACQUIRE program Grant No.~1640959), the Princeton Center for Complex Materials (PCCM), an NSF MRSEC (DMR-1420541), the Air Force Office of Scientific Research (Grant No.~FA9550-18-1-0081), the DARPA DRINQS program (Grant No.~D18AC00015), the Eric and Wendy Schmidt Transformative Technology Fund and the Princeton Catalysis Initiative. We acknowledge the use of Princeton's Imaging and Analysis Center, which is partially supported by PCCM, as well as the Princeton Micro-Nano Fabrication Lab and Quantum Device Nanofabrication Lab facilities. C.M.P.~was supported by the Department of Defense through the National Defense Science \& Engineering Graduate Fellowship Program.

\newpage
\begin{figure}[!h]
	\centering
	\includegraphics[width=87 mm]{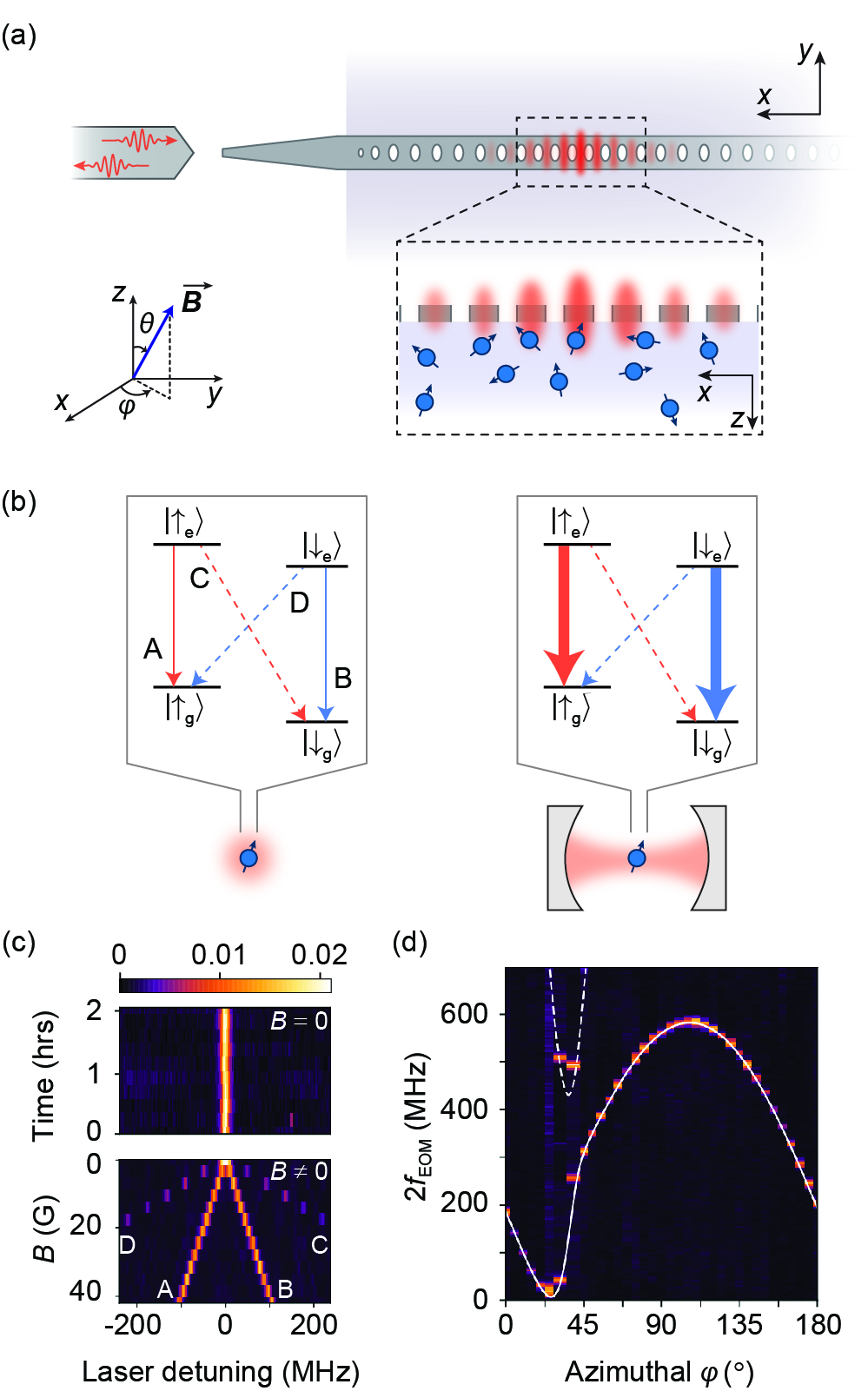}
	\caption{
	\textbf{Experimental approach.} 
	(\textbf{a}) The experimental device is a silicon photonic crystal cavity on top of an \er\!\!-doped YSO crystal. Light in the cavity evanescently couples to the \er ions. (inset) Definition of magnetic field angle $(\varphi,\theta)$; $(x,y,z)$ refer to the $(D_1,D_2,b)$ optical axes of the YSO crystal.
	(\textbf{b}) Without a cavity, the spin-conserving transitions A,B and spin-non-conserving transitions C,D are comparable in strength. The cavity selectively enhances A,B, resulting in highly cyclic optical transitions.
	(\textbf{c}) In the absence of a magnetic field, the four transitions are degenerate and give rise to a single, stable optical transition with a full width at half maximum of 6~MHz (centered at $\lambda = 1536.48$ nm). A magnetic field lifts the degeneracy. The color bar denotes the fluorescence intensity (arbitrary units).
	(\textbf{d}) The Zeeman splitting is strongly anisotropic, measured here by applying a 112 G magnetic field at various angles $\varphi$ ($\theta = 90\degree$) while driving the ion with a phase-modulated laser containing frequencies $f_0 \pm f_{EOM}$, where $f_0$ is the transition frequency when $B=0$. The solid (dashed) line shows the predicted splitting between the A-B (C-D) transitions  \cite{Sun:2008ik}.
	\label{fig:1}
	}
\end{figure}

\newpage
\begin{figure}[!h]
	\centering
	\includegraphics[width=170 mm]{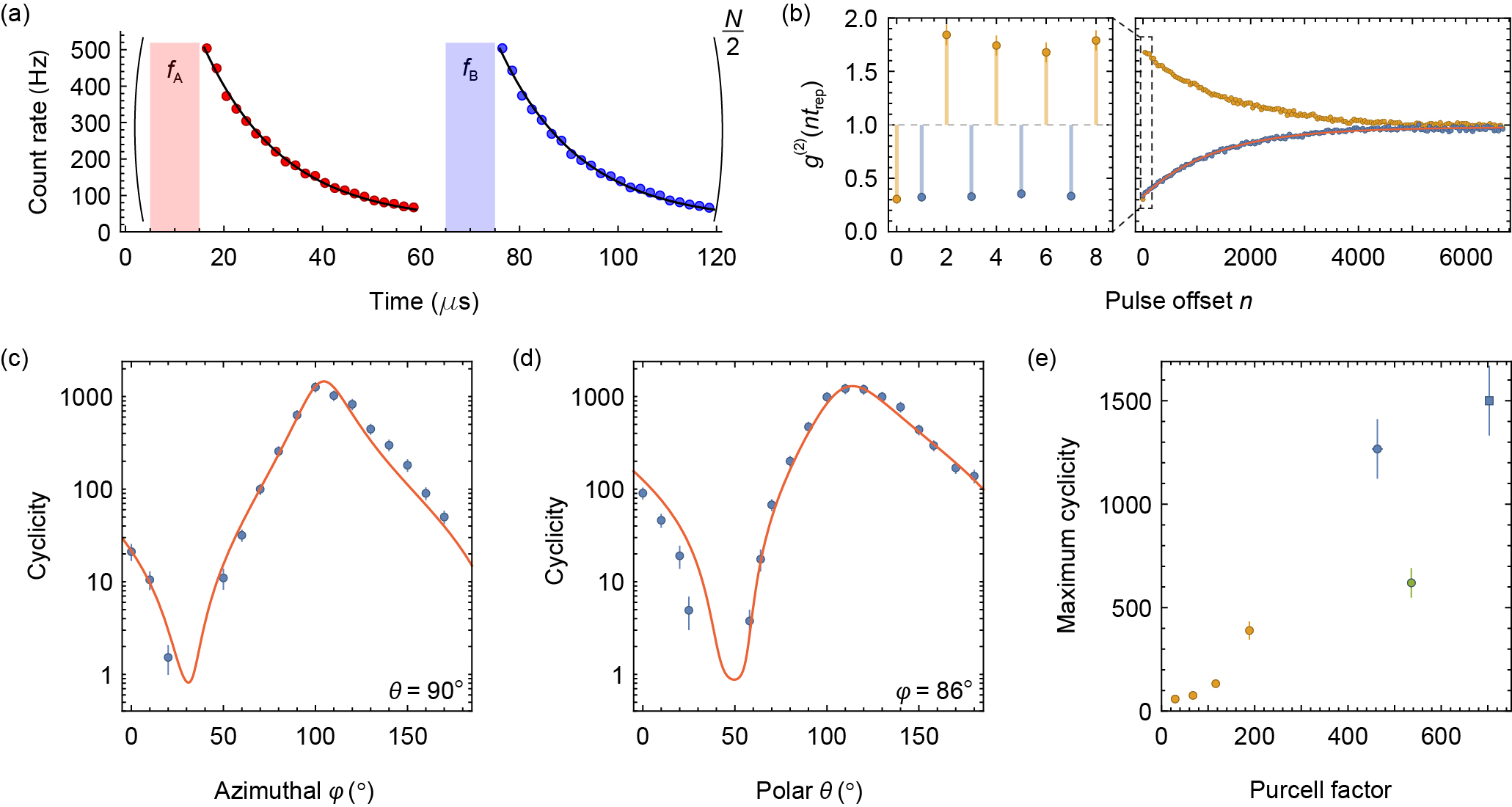}
	\caption{
	\textbf{Measuring the cyclicity of the optical transitions.}
	(\textbf{a}) The measurement sequence consists of alternating pulses on the A and B transitions (10 $\mu$s) followed by a fluorescence collection window (45 $\mu\text{s}$), repeating every $t_{rep} = 60 \mu$s. The fluorescence lifetime is $16.2 \pm 0.2\,\mu$s, shortened from its free-space value of 11.4 ms by $P=703\pm6$ in this device.
	(\textbf{b}) The intensity autocorrelation $g^{(2)}$ is computed from the integrated fluorescence after each pulse. Odd values of $n$ (blue points) probe the correlation between pulses driving different transitions and are anti-bunched as a result of the spin staying in the same state over many excitation cycles. $g^{(2)}$ decays as $e^{-n/n_0}$ because of optical pumping, giving the cyclicity $C \approx n_0/2$. In this measurement, {$C=660 \pm 66$}. $g^{(2)}(0) = 0.3$ is consistent with the signal-to-background ratio of 10.
	(\textbf{c}, \textbf{d}) The cyclicity varies dramatically with the angle of the external magnetic field, but is described by a model (red line) based on the independently measured $\mathbf{g}$ tensors (see text).
	(\textbf{e}) The cyclicity decreases rapidly with the Purcell factor, demonstrated here by measurements on several ions (shown in different colors) with the detuning varied to change $P$.
    }
	\label{fig:2}
\end{figure}

\newpage
\begin{figure}[!h]
	\centering
	\includegraphics[width=87 mm]{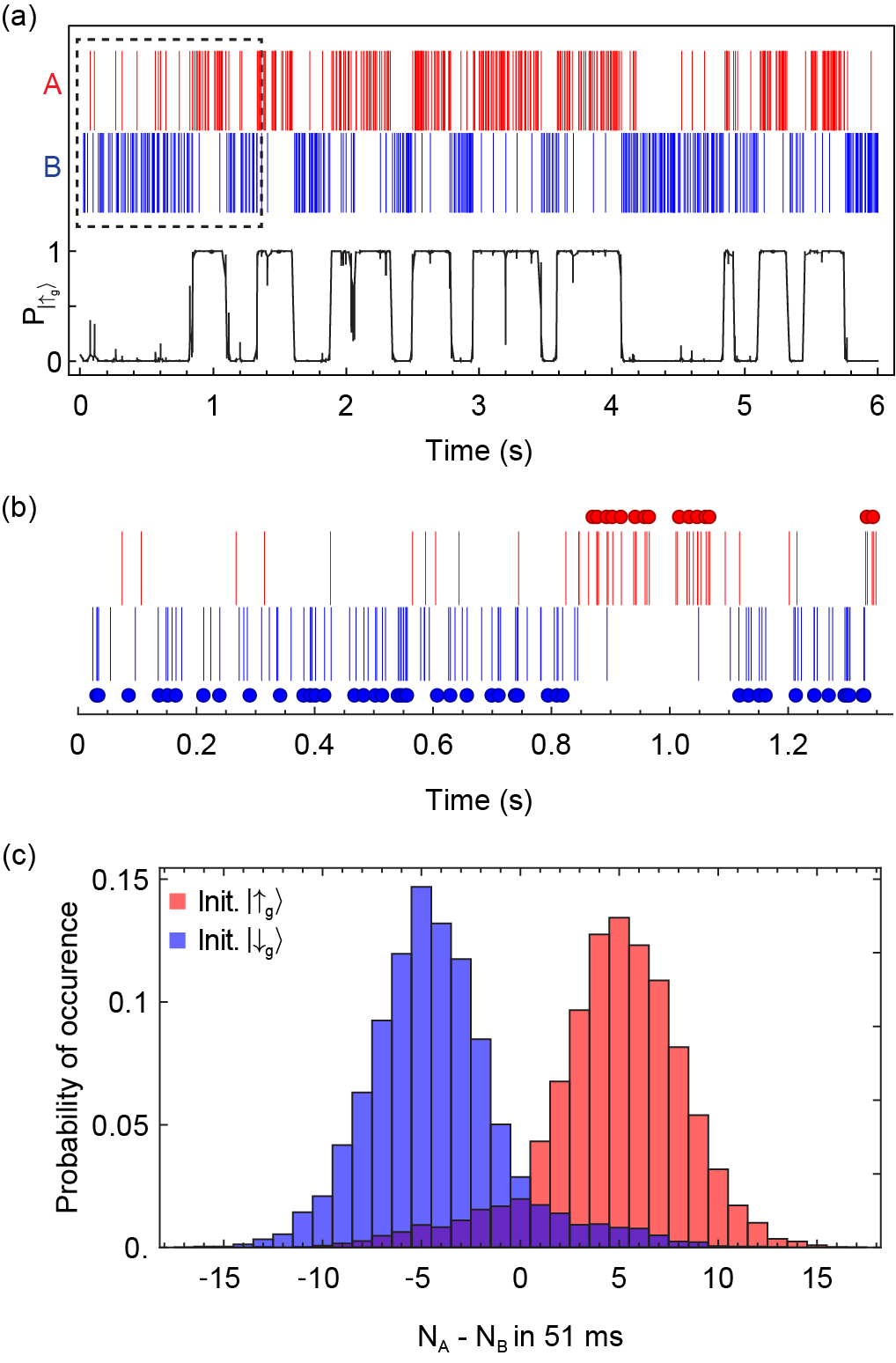}
	\caption{
	\textbf{Quantum nondemolition spin measurement.}
	(\textbf{a})
	Photons detected during a single run of the experiment using the sequence in Fig.~2a. The black curve shows the probability to be in $\ket{\upg}$ inferred from a Bayesian analysis, revealing quantum jumps between the spin states induced by optical pumping.
	(\textbf{b}) An adaptive maximum likelihood (ML) algorithm is used to perform successive single-shot spin measurements, with the result of each, independent measurement indicated by a circle [data from dashed box in (a)].
	(\textbf{c}) Histogram of photon counts for a fixed integration window of 51 ms following initialization using a ML measurement. The horizontal axis is the difference between the number of A and B photons detected in the measurement window.
	}
	\label{fig:3}
\end{figure}

\newpage
\begin{figure}[!h]
	\centering
	\includegraphics[width=170 mm]{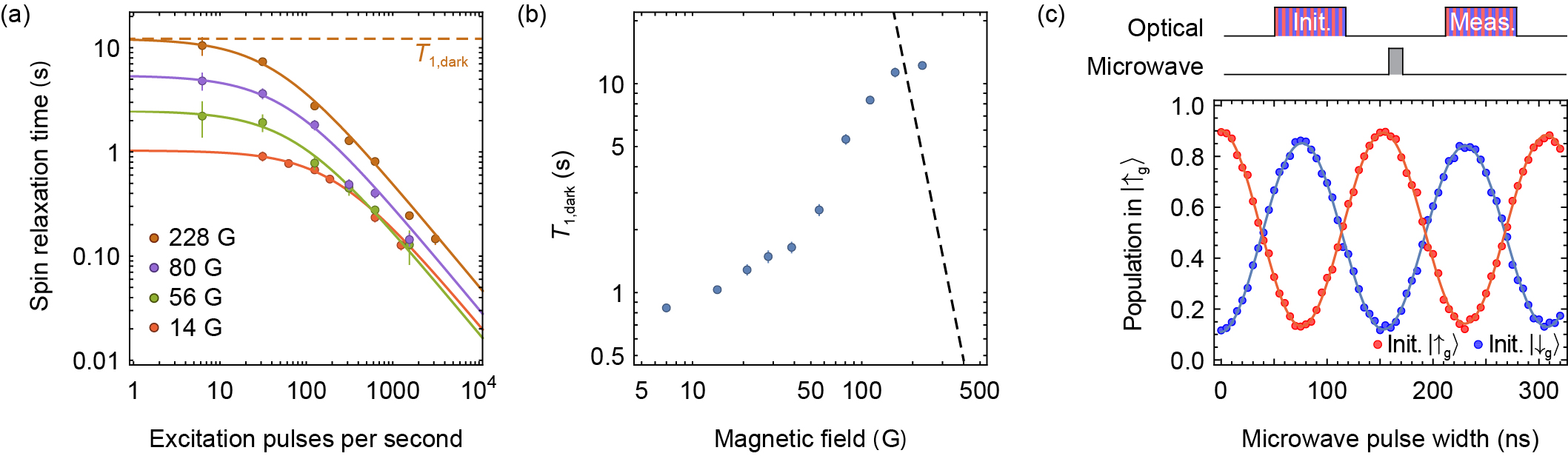}
	\caption{
	\textbf{Spin dynamics of a single \er ion.}
	(\textbf{a}) Spin relaxation times measured at varying repetition rates of the pulse sequence for several magnetic field amplitudes [$(\varphi,\theta)=(100,90)\degree$]. At low excitation rates, the spin relaxation time becomes independent of the optical excitation rate, revealing an intrinsic relaxation time $\td$.
	(\textbf{b}) $\td$ varies strongly with the amplitude of the magnetic field, achieving a maximal value of $12.2 \pm 0.4$ s. The dashed line indicates the expected spin-lattice relaxation rate \cite{Kurkin:1980jx, suppinfo}.
	(\textbf{c}) After initializing the spin with a projective measurement, Rabi oscillations can be observed between the two states using a microwave pulse of variable length ($f_\text{MW} = 1.76$~GHz for $B=112$~G). The oscillation contrast is consistent with a measurement fidelity of $\sim95\%$ for this ion (ion~3), which enters twice through the initial and final measurements.
	}
	\label{fig:4}
\end{figure}

\newpage
\bibliography{erbium_single_shot.bib}

\end{document}


\maketitle

\section{Experimental Details}

\begin{table}[t]
	\small
	\centering
    \begin{tabular}{|c || c | c | c |c| c | c| c|} 
	\hline
	\begin{tabular}{@{}c@{}} Ion \\ index \end{tabular} & 
	\begin{tabular}{@{}c@{}} Purcell \\ factor\end{tabular} & 
	\begin{tabular}{@{}c@{}} Maximum \\ cyclicity ($C_\text{max}$)\end{tabular} & 
	\begin{tabular}{@{}c@{}} Cavity \\ $Q$ factor\end{tabular} &
	\begin{tabular}{@{}c@{}} Cavity \\ losses ($\eta_\text{cav}$)\end{tabular} &
	\begin{tabular}{@{}c@{}} Total collection \\ efficiency ($\eta$) \end{tabular} &
	\begin{tabular}{@{}c@{}} ML readout\\ fidelity \end{tabular}
	\\
    \hline
    ion 1 (Fig.~3) & $703 \pm 6$ & {$1500 \pm 150$} & $6.6\times10^4$ & 0.063 & $ 0.028 $ & 0.946 \\
    ion 1 (Fig.~2) & $463 \pm 10$ & {$1260 \pm 126$} & $4.3\times10^4$ & 0.045 & $ 0.020 $ & - \\
    ion 2 & $189 \pm 5$ & {$390 \pm 39$} & $7.3\times10^4$ & 0.088 & $ 0.037 $ & 0.83 \\
    ion 3 & $536 \pm 5$ & {$620 \pm 62$} & $4.8\times10^4$ & 0.159 & $ 0.038 $ & 0.968 \\
	\hline
	\end{tabular}
    \caption{\textbf{Parameters for the 3 ions studied.} Cavity losses $\eta_\text{cav} = \kappa_\text{wg}/(\kappa_\text{wg} + \kappa_\text{int})$, where $\kappa_\text{wg}$ and $\kappa_\text{int}$ are waveguide and internal loss channels from the cavity, respectively, such that $\kappa_\text{wg} + \kappa_\text{int} =\kappa$ is the cavity linewidth. $\eta$ includes $\eta_\text{cav}$ as well as the measured fiber-waveguide coupling efficiency (typ. 40-60\%), other measured component losses and the detector quantum efficiency (67\%).
    }
    \label{tab:ion_cav_details}
\end{table}

\par The devices and substrates used in this work are identical to those in Ref.~\citealp{Dibos:2018hh}, and are described in detail there. The key difference in the present work is that the measurements take place in a $^3$He cryostat with a base temperature of approximately $T=540$ mK. At $T=4$ K, spin dynamics are unobservable, presumably because of rapid spin-lattice relaxation in the ground or excited states \cite{Welinski:2019jd}. The measurement of Rabi oscillations in Fig.~4c uses a slightly different device geometry that incorporates a microwave coplanar waveguide approximately 125 $\mu$m from the photonic crystal. Microwave pulses are generated using a signal generator (SRS SG386) modulated by an IQ mixer driven by an arbitrary waveform generator (Agilent 33622) and amplified to 21 W (MiniCircuits ZHL-30W-252+) before entering the cryostat. A low duty cycle is used to avoid heating the sample.

\par All data in this paper comes from measurements on 3 different ions (Table~\ref{tab:ion_cav_details}). Figs.~1,2,3, Figs.~\ref{fig:PexVsPhi} and \ref{fig:Ion1_darkt1_vs_phi_strength} are based on ``ion 1'', while Fig.~4a,b,   \ref{fig:cycVsDetuning}, \ref{fig:cycVsMag} and part of Fig.~2e, are based on ``ion 2''. Fig.~4c and part of  Fig.~2e are based on ``ion 3''. Ions 1 and~2 are coupled to different photonic crystals on the same YSO substrate, while ion 3 is in a different YSO crystal. Out of the many ions coupled to each cavity, these particular ions were selected for careful study because they are strongly coupled to the cavity (large Purcell factor) and spectrally well-separated from other ions. Importantly, no additional selection was made on the basis of cyclicity or spin readout fidelity. The transition from ion 1 to ion 2 was necessitated by the accidental destruction of the photonic crystal coupled to ion 1 (which was also damaged, lowering $Q$, after the measurements in Fig.~2a and 3 but before those in Fig.~2cd), while the transition to ion 3 was motivated by a new device geometry allowing for microwave driving of the spins.

\section{Theoretical model of cavity-enhanced cyclicity}

In this section, we develop a theoretical model describing the cyclicity of the optical transitions measured in Fig.~2 of the main text. Calculating the cavity coupling strength for all four possible transitions A-D is not currently possible because the relevant transition dipole moments of \er\!\!:YSO have not been measured, and the precise position of the ion in the cavity is not known. As an alternate approach, we demonstrate that these four rates and their dependence on the magnetic field angle can be reduced to only two fit parameters that physically correspond to the decay rates of the AB and CD transitions into the cavity at a single magnetic field orientation. The agreement of this model with the data validates our interpretation of the underlying physics, and is also practically useful for predicting conditions where the cyclicity is maximized from a small number of measurements.

\subsection{Atom-cavity Hamiltonian}
The Er$^{3+}$ $^4I_{13/2} \rightarrow ^4I_{15/2}$ transition in YSO has roughly equal contributions from electric (ED) and magnetic (MD) dipole transition amplitudes \cite{Dodson:2012dq}. The Hamiltonian describing this mixed coupling to a field is \cite{barron2009molecular}:
\begin{equation}
    H = -\vec{d}\cdot\vec{E} - \vec{\mu}\cdot\vec{B}
\end{equation}
Here, $\vec{d}$ is the electric dipole operator and $\vec{\mu}$ is the magnetic dipole operator. Typically, expressions for multipolar coupling include electric quadropole (E2) terms at the same order as MD contributions. However, for Er$^{3+}$, it has been calculated that the vacuum E2 emission rate is approximately 7 orders of magnitude smaller than the MD rate \cite{Dodson:2012dq}, justifying its exclusion in the present analysis.

We quantize the electric and magnetic fields in the cavity as:
\begin{align}
\vec{E} &=\, \vec{E}(r) a e^{-i \omega t} + \vec{E}^*(r) a^\dag e^{i \omega t} \\
\vec{B} &=\, i \left[\vec{B}(r) a e^{-i \omega t} - \vec{B}^*(r) a^\dag e^{i \omega t}\right]
\end{align}
The $i$ appearing in the expression for $\vec{B}$ reflects the fact that the magnetic field oscillates out of phase with the electric field in a standing wave cavity. For nanophotonic cavities, the TE/TM polarization mode splitting is high enough that we only consider a single polarization mode.

We then introduce four transition dipole operators $\sigma_i$, corresponding to the four transitions A-D. Each operator couples to electric and magnetic fields via the dipole moments $\vec{d}_i$ and $\vec{\mu}_i$, respectively. The complete atom-photon interaction Hamiltonian is  ($\hbar = 1$):

\begin{equation}
\label{eq:Hd0}
H_d = - \sum_{i=1}^4 
\left(\vec{d}_i \sigma_i + \vec{d}^*_i \sigma^{\dag}_i\right)\cdot \left(\vec{E}(\vec{r}_0)a e^{-i \omega t} + \vec{E}^*(\vec{r}_0)a^\dag e^{i \omega t}\right) 
{-} \,i \left(\vec{\mu}_i \sigma_i + \vec{\mu}^*_i \sigma^{\dag}_i\right) \cdot \left(\vec{B}(\vec{r}_0)a e^{-i \omega t} - \vec{B}^*(\vec{r}_0) a^\dag e^{-i \omega t}\right)
\end{equation}
Making the rotating wave approximation and taking the cavity to be resonant with the atomic transition (neglecting Zeeman splittings), we arrive at:
\begin{equation}
\label{eq:Hd1}
H_d = - \sum_{i=1}^4 \left(\vec{d}_i \cdot \vec{E}^*(\vec{r}_0) \sigma_i a^\dag - i \,\vec{\mu}_i \cdot \vec{B}^*(\vec{r}_0)\sigma_i a^\dag\right) + h.c.
\end{equation}
In the limit $\kappa \gg (\vec{d}\cdot\vec{E},\vec{\mu}\cdot\vec{B})$, the atom-cavity dynamics are simply Purcell-enhanced spontaneous emission into the cavity (the \emph{bad-cavity} limit of cavity QED). The decay rate on the transition $i$ into the cavity is given by:
\begin{equation}
    \Gamma_i = \frac{\left|\vec{d}_i \cdot \vec{E}^*(\vec{r}_0) - i \vec{\mu}_i \cdot \vec{B}^*(\vec{r}_0)\right|^2}{\kappa},
\end{equation}
where $\kappa$ is the cavity linewidth. In previous work, we have shown that the contributions from electric and magnetic coupling to the cavity could be of similar magnitude \cite{Dibos:2018hh}, depending on the position of the ion within in the cavity standing wave.

\subsection{Reducing parameters using Kramers' theorem}


\par Er$^{3+}$ has an odd number of electrons, so in the absence of a magnetic field, its eigenstates are all even-fold degenerate according to Kramers' theorem. In a low-symmetry environment like the Y site in YSO ($C_1$ point group), the degeneracy is minimal and all eigenstates are doublets \cite{abragam2012electron}. The application of a magnetic field lifts the degeneracy of the doublets, resulting in the singly degenerate states $\{\downg,\upg,\downe,\upe\}$ shown in Fig.~1b.

The states emerging from the same doublet are time-reversal conjugates of each other: $\hat{\Theta} \ket{\uparrow_j} = \ket{\downarrow_j}$ and $\hat{\Theta} \ket{\downarrow_j} = -\ket{\uparrow_j}$, where $\hat{\Theta}$ is the antiunitary time-reversal operator. This has implications for the matrix elements of the electric and magnetic dipole operators, which are even and odd under time reversal ($\Theta^{-1}A\hat{\Theta} = \pm A$), respectively \cite{abragam2012electron}. In particular:

\begin{eqnarray}
\label{eq:timereversal1}
d_1 &=& \bra{\downe}d\ket{\downg} = \bra{\hat{\Theta}\upe}d\ket{\hat{\Theta}\upg} = \bra{\upe}\hat{\Theta}^{-1}d\hat{\Theta}\ket{\upg}^* = \bra{\upe}d\ket{\upg}^* = d_4^* \\
d_2 &=& \bra{\downe}d\ket{\upg} = -\bra{\hat{\Theta}\upe}d\ket{\hat{\Theta}\downg} = -\bra{\upe}\hat{\Theta}^{-1}d\hat{\Theta}\ket{\downg}^* = -d_3^* \\
\mu_1 &=& -\mu_4^* \\
\mu_2 &=& \mu_3^*
\label{eq:timereversal4}
\end{eqnarray}
We can now revisit the atom-photon coupling Hamiltonian, Eqn.~(\ref{eq:Hd1}). Writing the atomic operators $\sigma_i$ in the basis ${\{\downg,\upg,\downe,\upe\}}$:

\begin{eqnarray}
H_d &=& - 
\left(
\begin{array}{cccc}
     0 & 0 & \vec{d}_1 & \vec{d}_3  \\
     0 & 0 & \vec{d}_2 & \vec{d}_4 \\
     0 & 0 & 0 & 0  \\
     0 & 0 & 0 & 0
\end{array}
\right) \cdot \vec{E}^{{*}}(\vec{r_0}) a^\dag + i 
\left(
\begin{array}{cccc}
     0 & 0 & \vec{\mu}_1 & \vec{\mu}_3  \\
     0 & 0 & \vec{\mu}_2 & \vec{\mu}_4 \\
     0 & 0 & 0 & 0  \\
     0 & 0 & 0 & 0
\end{array}
\right) \cdot \vec{B}^{{*}}(\vec{r_0}) a^\dag + h.c. \\
&=& - 
\left(
\begin{array}{cccc}
     0 & 0 & \vec{d}_{||} & \vec{d}_{\perp}  \\
     0 & 0 & -\vec{d}_{\perp}^* & \vec{d}_{||}^* \\
     0 & 0 & 0 & 0  \\
     0 & 0 & 0 & 0
\end{array}
\right) \cdot \vec{E}^{{*}}(\vec{r_0}) a^\dag + i 
\left(
\begin{array}{cccc}
     0 & 0 & \vec{\mu}_{||} & \vec{\mu}_{\perp}  \\
     0 & 0 & \vec{\mu}_\perp^* & -\vec{\mu}_{||}^* \\
     0 & 0 & 0 & 0  \\
     0 & 0 & 0 & 0
\end{array}
\right) \cdot \vec{B}^{{*}}(\vec{r_0}) a^\dag + h.c. \\
&=& - 
\left(
\begin{array}{cccc}
     0 & 0 & g^e_{||} - i g^m_{||} & g^e_{\perp} - i g^m_{\perp} \\
     0 & 0 & -g^{e\,*}_{\perp} - i g^{m\,*}_{\perp} & g^{e\,*}_{||} + i g^{m\,*}_{||} \\
     0 & 0 & 0 & 0  \\
     0 & 0 & 0 & 0
\end{array}
\right) a^\dag + h.c. \\
&=& -  
\left(
\begin{array}{cccc}
     0 & 0 & g_{||} & g_{\perp} \\
     0 & 0 & -g_{\perp}^* & g_{||}^* \\
     0 & 0 & 0 & 0  \\
     0 & 0 & 0 & 0
\end{array}
\right) a^\dag + h.c.
\label{eq:hacfinal}
\end{eqnarray}
In going from the first to the second line, we have made use of Eqns.~(\ref{eq:timereversal1}) - (\ref{eq:timereversal4}). In going from the second to the third line, we have taken the cavity fields to be real-valued, which is an excellent approximation as they are highly linearly polarized.

The final expression, Eqn.~(\ref{eq:hacfinal}), has the same form for the electric and magnetic dipole moments despite the difference in their transformation properties in Eqns. (\ref{eq:timereversal1}) - (\ref{eq:timereversal4}). Mathematically, this arises from the factor of $i$ in front of the magnetic field term in Eqn.~(\ref{eq:Hd1}). Physically, this means that interference of the electric and magnetic dipole decay channels into the cavity does not introduce any chirality (\emph{i.e.}, preference for decays from $\ket{\upe}$ relative to $\ket{\downe}$) into the atom-cavity system, which is intuitive for a standing-wave optical cavity.


\subsection{Angle dependence of the cyclicity}
We now turn to computing the cyclicity and its dependence on the orientation of the external magnetic field defining the spin quantization axis. The branching ratio between spin-non-conserving and spin-conserving decays through the cavity is 
{$R=\Gamma_{\perp}/(\Gamma_{||}+\Gamma_{\perp}) = |g_{\perp}|^2/(|g_{||}|^2 + |g_{\perp}|^2)$}, and the cyclicity is ${C = 1/R}$.
The spin eigenstates are given by the effective spin Hamiltonian:
\begin{equation}
\label{eq:Hz}
    H_{Z} = \mu_B \vec{B} \cdot \mathbf{g} \cdot \vec{S},
\end{equation}
where $\vec{B}$ is the applied magnetic field, $\vec{S} = \{\sigma_x,\sigma_y,\sigma_z\}$ is a vector of Pauli matrices, $\mu_B$ is the Bohr magneton and $\mathbf{g}$ is a symmetric, real 3x3 matrix. For Er$^{3+}$:YSO, $\mathbf{g}$ is highly anisotropic in both the ground and excited electronic states, with principal components $(14.65,1.80,0.56)$ for the ground state and $(12.97,0.85,0.25)$ for the excited state~\cite{Sun:2008ik}. Note that the orientation of the eigenvectors of $\mathbf{g}$ is different from the crystal axes and also slightly different between the ground and excited states. A consequence of the anisotropy of $\mathbf{g}$ is that the spin eigenvectors are not generally parallel to the applied magnetic field.

The eigenvectors at one magnetic field orientation $(\varphi',\theta')$ can be expressed in terms of those at another orientation as: $\ket{\upg(\varphi',\theta')} = \alpha \ket{\upg(\varphi,\theta)} + \beta \ket{\downg(\varphi,\theta)}$. Denoting the matrix of atom-cavity coupling matrix elements in Eqn.~(\ref{eq:hacfinal}) as $M(\varphi,\theta)$, we can transform it to another basis according to:
\begin{eqnarray}
    g_{||}(\varphi',\theta') &=& \bra{\downg(\varphi',\theta')}M(\varphi,\theta)\ket{\downe(\varphi',\theta')} \\
    g_{\perp}(\varphi',\theta') &=& \bra{\downg(\varphi',\theta')}M(\varphi,\theta)\ket{\upe(\varphi',\theta')} \\
\end{eqnarray}

and evaluate the cyclicity as before.

We note that a similar model was applied to the orientation dependence of the \emph{absorption} of linearly polarized light by Nd$^{3+}$:YVO$_4$ in Ref.~\citealp{Afzelius:2010gb}. The applicability of this model to \emph{emission} in our work stems from the restriction of the emission to a single polarization by the high Purcell factor coupling to a single-mode, non-degenerate cavity.

This model is used to fit the data in Fig. 2 of the main text. The model is fit using $M(100 ^{\degree}, 90^{\degree})$ as the fit parameters, and the result is $g_{||} = e^{-i 1.15}$, $g_{\perp} = 0.024 \times e^{-i 1.476}$. Since the cyclicity only depends on the ratio of these quantities, we constrain $|g_{||}|=1$.

\subsection{Correction for free-space decay}
As the Purcell factor is finite, free-space emission can influence the cyclicity when it is very large. We incorporate this by adding free space decays to the branching ratio expression, such that the cyclicity becomes $C_{0}$ when the cavity coupling vanishes:

\begin{equation}
\label{eq:cycwfreespace}
    C = \frac{\Gamma_{CD} + \Gamma_{AB}}{\Gamma_{CD}} = 1 + \frac{\Gamma_{AB}^0 + |g_{||}|^2/\kappa}{\Gamma_{CD}^0 + |g_{\perp}|^2/\kappa} = 1+\frac{1-1/C_0 + P_{||}}{1/C_0 + P_{\perp}}
\end{equation}

In this expression, $\Gamma_i^0$ denotes the decay rate on transition $i$ in the absence of a cavity, $C_0 = 1 + \Gamma_{AB}^0/\Gamma_{CD}^0$ is the cyclicity in the absence of a cavity, and $P_{||/\perp} = g_{||/\perp}^2/(\kappa \Gamma^0)$, where $\Gamma^0$ is the total decay rate out of the excited state in the absence of the cavity.

For the ions studied here with $P$ of order several hundred, the inclusion of this correction does not meaningfully improve the fit. We believe this occurs because the fit function can artificially increase $g_{\perp}$ by a small amount at the orientation of maximum cyclicity to account for the free space decay without significantly impacting the cyclicity at other angles. However, measuring the same ion with different Purcell factors (realized by changing the cavity detuning) makes the free-space decay evident and allows $C_0$ to be estimated (section \ref{sec:cbare}).

\subsection{Highest achievable cyclicity}
For an atom with Purcell factor $P$ and bare cyclicity $C_0$, the highest possible cyclicity is $C = P C_0$, achieved when $g_{\perp} = 0$. This can always be realized by choosing the cavity polarization to be perpendicular to $\vec{d}_{\perp}$, for some spin quantization axis. In contrast, if the cavity polarization is fixed, it may be possible to tune the quantization axis angle to achieve $g_{\perp} = 0$. We consider three cases:

\begin{enumerate}
    \item If the $\mathbf{g}$ tensors in the ground and excited state are isotropic (or equal), it is always possible to choose an orientation of $\vec{B}$ where $g_{\perp} = 0$. This can be proved from the normality of the matrix $m$ formed by the upper-right 2x2 block of $M$.
    \item If the $\mathbf{g}$ tensors for the ground and excited states are not simultaneously diagonalizable (having different axes), then it is not possible to achieve $g_{\perp} = 0$ in general. This is proved by the specific counterexample of \er\!\!:YSO, where we observe (in a numeric model) that $g_{\perp}$ can be made small but never quite zero for certain values of $M$.
    \item If the $\mathbf{g}$ tensors for the ground and excited states are parallel but have different (non-zero) magnetic moments, it appears (numerically) to always be possible to make $g_{\perp} = 0$. We have not proven this mathematically.
\end{enumerate}

\er\!\!:YSO falls into the second case because of the $C_1$ site symmetry of the Er site. However, the principal axes of the ground and excited state $\mathbf{g}$ tensors are only rotated by about $15\degree$ from each other \cite{Sun:2008ik}, which may explain the success in achieving high cyclicity anyway. Many other host crystals for REIs and other defects have higher site symmetry, which forces the alignment of the g tensor axes to the crystal, and are therefore covered by the third case. Therefore, we expect that this technique is fairly general.


\section{Calibration of $P_\text{ex}$}

\begin{figure}[!h]
	\centering
	\includegraphics[]{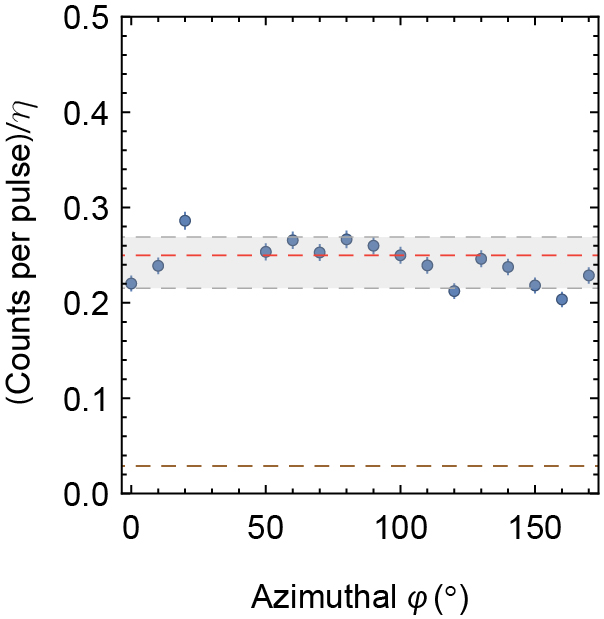}
	\caption{
	\textbf{$P_\text{ex}$ calibration}. Average number of collected photons following each excitation pulse during the measurement in Fig.~2c. The photon numbers are scaled by the independently measured collection and detection efficiency $\eta = 0.020$. Since the spin is on average unpolarized and only resonant with the laser half of the time, $P_\text{ex}=1/2$ (as defined here) corresponds to $1/4$ photon per pulse.
	}
	\label{fig:PexVsPhi}
\end{figure}

For the measurements in Fig.~2 of the main text, we use saturating optical pulses to ensure $P_\text{ex} \approx 0.5$ regardless of the magnetic field orientation. The approximate value of $P_\text{ex}$ is confirmed by counting the emitted photons and comparing to the independently measured collection and detection efficiency~(Fig.~\ref{fig:PexVsPhi}). We note that the finite duration of the excitation pulse may allow for $P_\text{ex} > 0.5$ if a photon is emitted during the pulse and the ion is re-excited, and have observed increased values of $C$ at certain angles when using optical $\pi$ pulses instead. In that sense, the values of $C$ that we quote should be interpreted as lower bounds.

\section{Estimate of intrinsic cyclicity of Er$^{3+}$:YSO transition}
\label{sec:cbare}

A central claim of our work is that the cyclicity of the ion is enhanced by the optical cavity. Since the cyclicity $C_0$ of the spin transitions in Er$^{3+}$:YSO without a cavity has not been previously measured, there is no direct basis for comparison. In ensemble experiments, $C_0$ has been estimated to be around 10 using measurements of the optical pumping rate and a number of simplifying assumptions \cite{HastingsSimon:2008fj}. We have attempted to measure $C_0$ using a single ion by detuning the cavity as much as possible, which increases the fraction of decays into free space and yields a weighted average of the cavity-enhanced $C$ and $C_0$ [Eq. \ref{eq:cycwfreespace}].

\begin{figure}[!htb]
	\centering
	\includegraphics[]{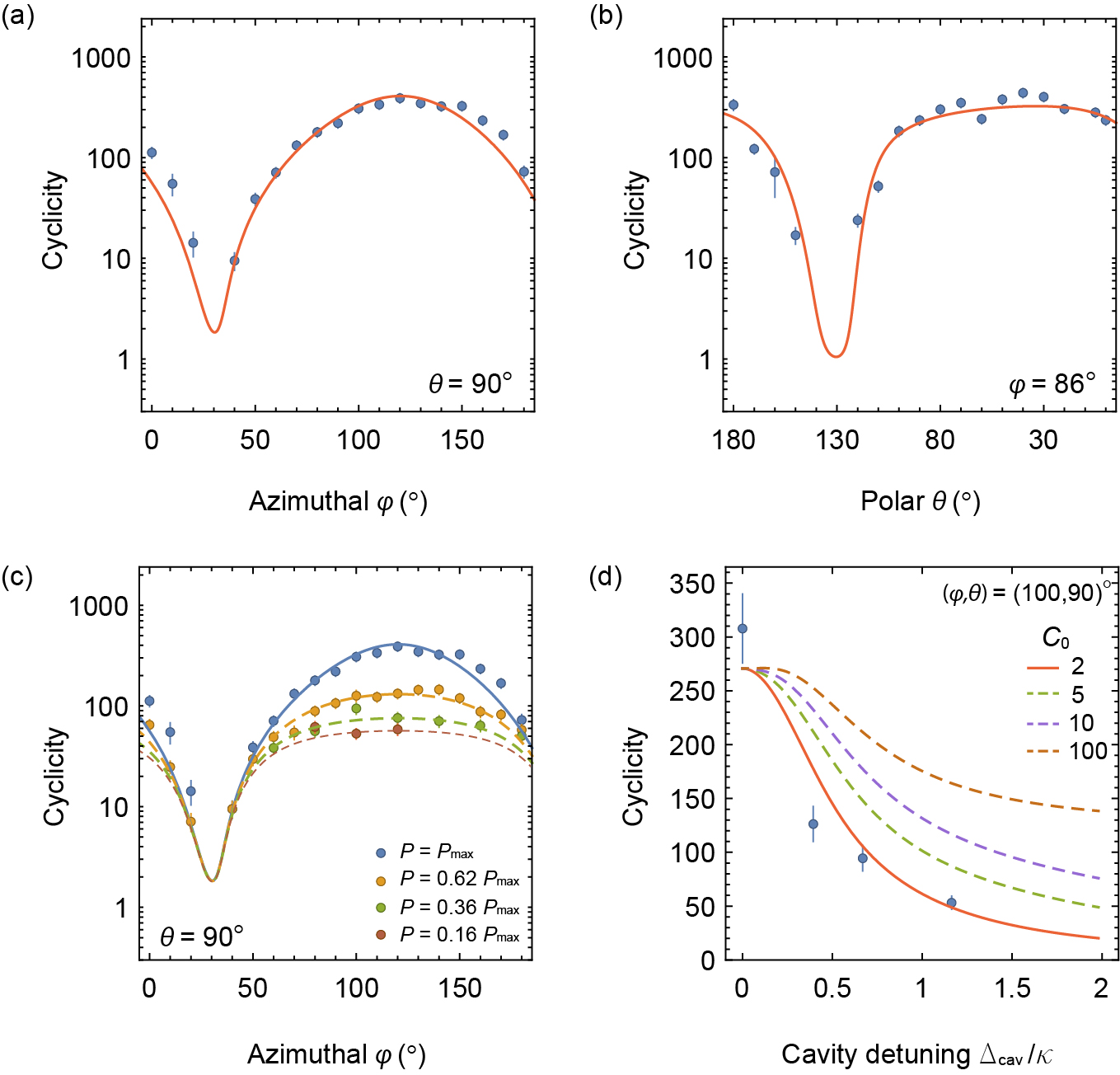}
	\caption{
	\textbf{Extracting the bare ion cyclicity, $C_0$}.
	(\textbf{a},\textbf{b}) Cyclicity measurements on ion 2 (situated in a different cavity) show a similar orientation dependence to ion 1. Each of the two crystallographically inequivalent Y sites in YSO has two possible orientations related by a $C_2$ rotation about the $b$ axis. Ion 2 is in the opposite site from ion 1, so we have inverted the $\theta$ axis to make the apparent dependence the same.
	(\textbf{c}) Measurement of cyclicity for the same ion 2 at several different Purcell factors, achieved by detuning the cavity from the atomic transition by an amount $\Delta_\text{cav}$. The data is fit to Eq. \ref{eq:cycwfreespace}. Here, $|B|=112$ G and $\theta = 90^{\degree}$.
	(\textbf{d}) Cyclicity at $(\varphi, \theta) =(100, 90)\degree$ vs. cavity detuning. The model is Eq. \ref{eq:cycwfreespace}, where $P_{||}$ and $P_{\perp}$ acquire a detuning dependence $P(\Delta_i) = \frac{P_\text{max}}{1+(2\Delta_i/\kappa)^2}$ based on the known Zeeman splittings of the four transitions. Since the decay rates out of the two excited states are no longer equal when the cavity is detuned (\emph{i.e.}, $\Gamma_A \neq \Gamma_B$), Eq. \ref{eq:cycwfreespace} is averaged over the two excited states. A fit yields $C_0 = 2 \pm 3.$
	}
	\label{fig:cycVsDetuning}
\end{figure}

In Fig.~\ref{fig:cycVsDetuning}{c}, we show several measurements of the same ion with increasing cavity detuning to decrease the total Purcell factor. The maximum cyclicity is strongly reduced: at the highest detuning, we observe a cyclicity of {$53 \pm 5$}, which sets an upper bound on $C_0$. In Fig.~\ref{fig:cycVsDetuning}d, we plot the cyclicity vs.~detuning together with a theoretical model for several values of $C_0$. From this model, we conclude that $C_0$ is less than {10}, and likely closer to {2}. Direct measurements at higher detunings are not possible because the count rate falls dramatically.

\section{Single-shot measurement fidelity}

Two factors contribute to the ML fidelity: the statistical error set by the signal-to-background ratio during the collection period (\textit{SNR}), and the probability to decay to the wrong state before enough photons are collected, $m/(\eta C)$, where $m$ is the number of photons needed to reach the target fidelity $F_\text{T}$. $m$ is related to the \textit{SNR} as $F_\text{T} = \textit{SNR}^m/(\textit{SNR}^m+1)$. In our experiment, \textit{SNR} is typically high enough (14 for ion 1, limited by a timing error, and around 20 for ions 2 and 3, limited by dark counts) that by the time a single photon is detected, the error from the finite cyclicity is larger than the statistical error. In this regime, a nearly-optimal strategy is to infer that the spin state is $\ket{\upg}$ whenever an ``A'' photon is detected, and $\ket{\downg}$ whenever an ``B'' photon is detected. The average measurement fidelity is $1-1/(\eta C)$, and the average measurement duration is $t_\text{meas} = 1/(P_\text{ex} \eta C)$.

Based on this model, we can project how improved devices might lead to improved measurements. Assuming the same demonstrated cyclicity ({$1500$}), fiber-waveguide coupling and photon detection efficiency, but improving the cavity to realize $Q_\text{int} = 10^6$ with critical coupling ($\kappa_\text{wg} = \kappa_\text{int}$) and using optical $\pi$ pulses to increase $P_\text{ex}$ to {1}, it should be possible to realize an average measurement fidelity of 0.996 in average time of 50 $\mu$s. This assumes that \textit{SNR} remains dark-count limited as the Purcell factor is increased.

\section{Intrinsic spin relaxation}

\begin{figure}[]
	\centering
	\includegraphics[]{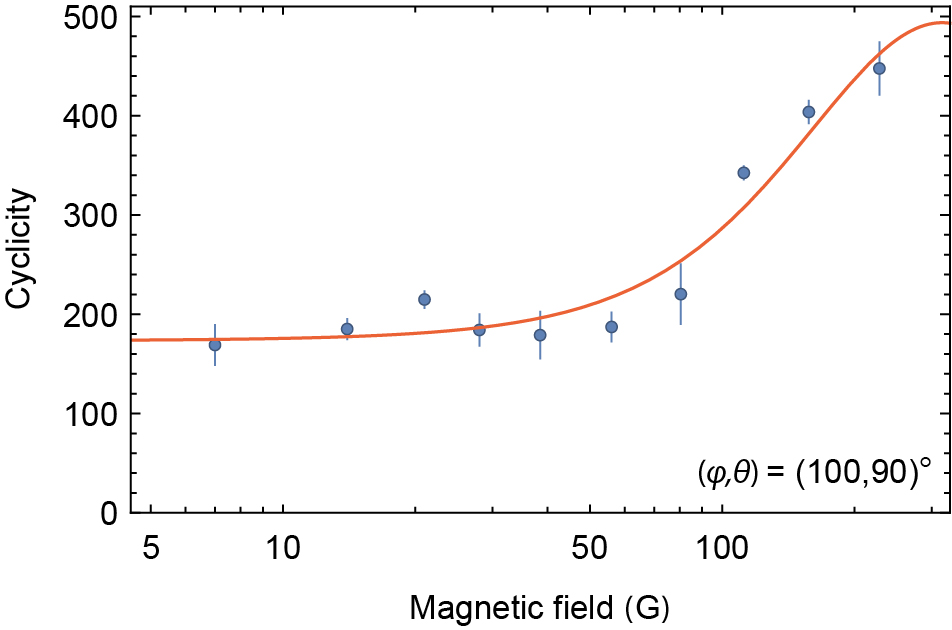}
	\caption{
	\textbf{Magnetic field amplitude dependence of cyclicity}. Based on the data for Ion~2 in Fig.~4a,b of the main text, we extract the cyclicity of the optical transitions at a fixed magnetic field orientation for varying field amplitudes. The data is fit to the model in Eq. \ref{eq:cycwfreespace}, incorporating the detuning-dependence of the Purcell enhancement as described in the caption to Fig.~\ref{fig:cycVsDetuning}. $C_0$ fits here to $5 \pm 1$. 
	}
	\label{fig:cycVsMag}
\end{figure}

The spin relaxation rates in Fig. 4a are fit to a model of the form $T_{SR}^{-1} = T_\text{1,dark}^{-1} + {P_\text{ex}/(t_\text{rep}C)}$, where the repetition rate $t_\text{rep}$ is varied to change the optical pumping strength. The intrinsic relaxation rate $T_\text{1,dark}$ varies strongly with the magnetic field amplitude, and the cyclicity $C$ also has a weak dependence. The latter is explained by the larger Zeeman shift of the spin-non-conserving transitions CD ($\sim 13.1$ MHz/Gauss) compared to the AB transitions ($\sim 2.5$ MHz/Gauss) shifting the former out of resonance with the cavity more quickly (Fig.~\ref{fig:cycVsMag}). From this data, we can also affirm that the selective Purcell enhancement of the spin-conserving transition does not primarily arise from detuning the CD transitions, as in Ref.~\citealp{Sun:2018hx}, although this effect does provide an additional factor of 2-3 at the highest magnetic fields.

\begin{figure}[]
	\centering
	\includegraphics[width=170 mm]{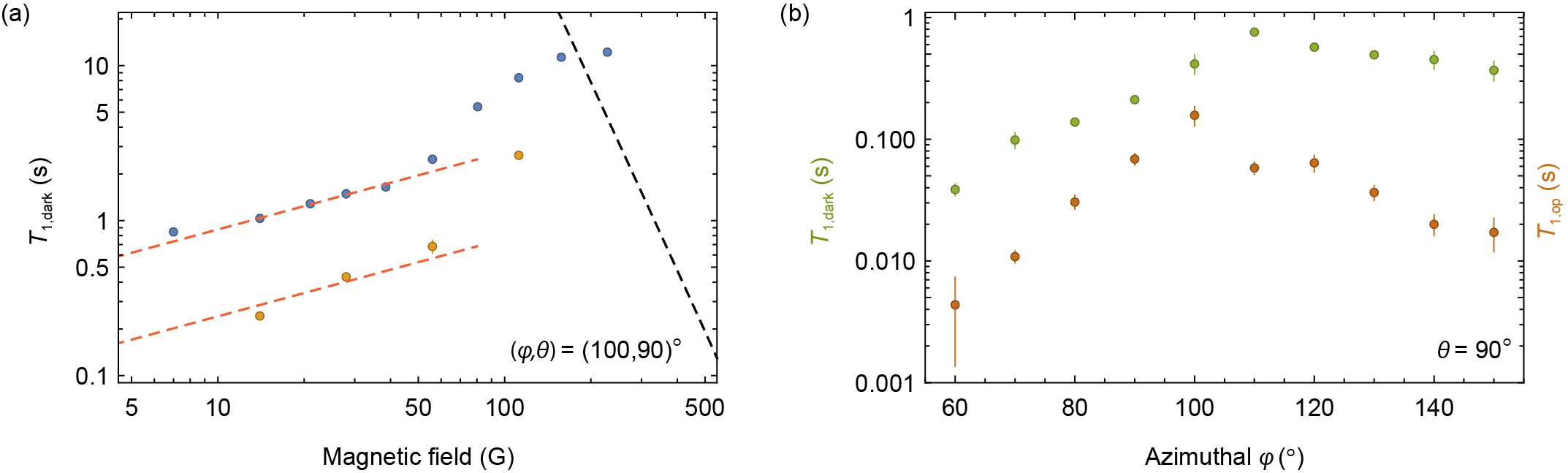}
	\caption{
	\textbf{Additional measurements of $T_\text{1,dark}$}. 
	(\textbf{a}) Intrinsic spin relaxation time ($T_\text{1,dark}$) of ion~1 (yellow) measured under the same conditions as ion~2 (blue points, reproduced from main text). The empirical $B^{1/2}$ scaling observed at low fields is indicated by the red, dashed line. The relaxation rate for ion~1 follows a similar trend to ion~2 but is 4 times faster. The black, dashed line is the model from Ref.  \citealp{Kurkin:1980jx}.
	(\textbf{b}) $T_\text{1,dark}$ (green points) measured at a few field orientations ($B=28$ G) reveal the anisotropic nature of $T_\text{1,dark}$. The angle dependence qualitatively agrees with flip-flop limited Er$^{3+}$:YSO lifetimes measured in Ref. \citealp{car2018optical}. The brown points denote the optical pumping time constants under these conditions, showing that the cyclicity measurements are generally not limited by $T_\text{1,dark}$.
	}
	\label{fig:Ion1_darkt1_vs_phi_strength}
\end{figure}

The magnetic field dependence of $T_\text{1,dark}$ disagrees markedly with predictions based on the measured coefficients for the Raman, Orbach and direct processes for \er\!\!:YSO \cite{Kurkin:1980jx}. At $T=0.5$ K, these are dominated by the direct process, which should have a magnetic field dependence $T_1 \propto B^{-4}$, which arises from the combination of the frequency-dependence of the phonon density of states and the magnetic field-dependence of the spin-phonon coupling~\cite{abragam2012electron}. The measured values display a $B^{1/2}$ dependence at low fields, transitioning to a more rapid increase around $B=50$ Gauss before saturating at 200 Gauss. We note that the point where $T_\text{1,dark}$ saturates is roughly consistent with the onset of the direct process, and that the disagreement could be attributed to anisotropy in the direct process rate. Anomalous magnetic field dependence in the low-field relaxation of rare earth ions has been previously observed in electron paramagnetic resonance \cite{Sabisky:1970gi} without a definitive explanation. Concentration-dependent spin relaxation has been observed in spectral hole burning experiments \cite{car2018optical} and attributed to flip-flop interactions between nearby Er ions. This is a likely explanation for our observations, which is bolstered by the difference in $T_\text{1,dark}$ for different ions [which may arise from the stochastic arrangement of ions in this low-density sample ([Er] $\approx 300$ ppb)] and a strong ansiotropy in  $T_\text{1,dark}$ (Fig.~\ref{fig:Ion1_darkt1_vs_phi_strength}).

\section{Additional data analysis}
\subsection{Correction of cyclicity estimate at small $n_0$ values}

In Fig.~2b and thereafter, the intensity autocorrelation $g^{(2)}$ has been fitted to an exponential function and the decay constant $n_0$ is utilized to extract $C = P_\text{ex}n_0$. This expression for $C$ is only valid when $n_0\gg 1$ because of the discrete time steps in the measurement. We estimate the value of $C$ more accurately by using the following expression:
\begin{equation}
    C = \frac{P_\text{ex}}{\left( 1- e^{-1/n_0} \right)},
    \label{eq:define_C}
\end{equation}
which reduces to the previous expression for $C$ at large $n_0$ values.

\subsection{Extracting $n_0$ from $g^{(2)}$ at low optical pumping rates}

The spin relaxation time is extracted from fits to $g^{(2)}$ measurements as described in Fig. 2b. When these time constants are longer than 1 s, the even- and odd-offset $g^{(2)}$ time constants begin to differ. We believe this results from spectral diffusion of the optical transitions. Under the assumption that this is uncorrelated with the spin dynamics and acts identically on the A and B transitions, we find that the spin time constant can be isolated by fitting an exponential to the \emph{difference} of the even- and odd-offset traces. 

\newpage
\bibliography{single_shot_SI.bib}